\documentclass[%
 reprint, 
 amsmath,amssymb,
 aps,
citeautoscript,
prb,
floatfix]{revtex4-1}
\usepackage{graphicx}%
\usepackage{dcolumn}%
\usepackage{bm}%
\usepackage{xcolor}
\usepackage[normalem]{ulem}
\usepackage{color,soul}
\usepackage[]{lineno}
\usepackage{rotating} 
\usepackage{booktabs} 
\usepackage{amsmath} 

\newcommand{\equalcontrib}{\textsuperscript{\dag}}

\begin{document}

\preprint{APS/123-QED}

\title{Low-power integrated optical parametric amplification via second-harmonic resonance}

\author{Devin~J.~Dean\textsuperscript{1}\equalcontrib}
\author{Taewon~Park\textsuperscript{1,2}\equalcontrib}
\author{Hubert~S.~Stokowski\textsuperscript{1}}
\author{Luke~Qi\textsuperscript{1}}
\author{Sam~Robison\textsuperscript{1}}
\author{Alexander~Y.~Hwang\textsuperscript{1}}
\author{Jason~Herrmann\textsuperscript{1}}
\author{Martin~M.~Fejer\textsuperscript{1}}
\author{Amir~H.~Safavi-Naeini\textsuperscript{1,{$\star$}}\vspace*{3 mm}}

\affiliation{%
\textsuperscript{1} \mbox{Department of Applied Physics and Ginzton Laboratory, Stanford University, Stanford, California 94305, USA}
}%
\affiliation{%
\textsuperscript{2} \mbox{Department of Electrical Engineering, Stanford University, Stanford, California 94305, USA}
}%
\affiliation{%
\textsuperscript{{$\star$} } safavi@stanford.edu \vspace*{-6 mm}
}%

\pacs{Valid PACS appear here}
\maketitle

{\bf
Optical amplifiers are fundamental to modern photonics, enabling long-distance communications, precision sensing, and quantum information processing. Erbium-doped amplifiers dominate telecommunications but are restricted to specific wavelength bands, while semiconductor amplifiers offer broader coverage but suffer from high noise and nonlinear distortions. Optical parametric amplifiers (OPAs) promise broadband, quantum-limited amplification across arbitrary wavelengths. 
However, their miniaturization and deployment has been hampered by watt-level power requirements. 
Here we demonstrate an integrated OPA on thin-film lithium niobate that achieves $>17$~dB gain with $<200$~mW input power -- an order of magnitude improvement over previous demonstrations. Our second-harmonic-resonant design enhances both pump generation efficiency ($95\%$ conversion) and pump power utilization through recirculation, without sacrificing bandwidth. The resonant architecture increases the effective pump power by nearly an order of magnitude compared to conventional single-pass designs, while also multiplexing the signal and pump. We demonstrate flat near-quantum-limited noise performance over $110$~nm. Our low-power architecture enables practical on-chip OPAs for next generation quantum and classical photonics.

}

\maketitle


\section{Introduction}\label{sec_intro}

Optical amplification underpins modern photonics, from transcontinental communications to quantum computing. Traditional amplifiers based on rare-earth dopants or semiconductors are limited by their fixed gain spectra and fundamental noise constraints. Optical parametric amplifiers (OPAs) overcome these limitations by using nonlinear interactions to transfer energy from a pump beam to signal photons~\cite{Jankowski_2021}. This makes them promising for extending fiber-optic communications beyond current bandwidth limits, e.g., those of erbium-doped fiber amplifiers \cite{Ho2001, Kobayashi:23, Shimizu2023, Shimizu2024, Kuznetsov2025}, as well as for spectroscopic sensing, where parametric amplifiers are uniquely able to generate and perform quantum-limited amplification of signals while extending to new wavelength ranges. OPAs can also provide phase-sensitive amplification enabling schemes such as nonlinearity-compensation in fiber networks \cite{Yariv:79, Umeki:16, Foo:19, Shimizu2024}, noiseless amplification of optical signals \cite{Kazama:21, Ye:2021}, and the generation of squeezed light for applications in quantum sensing and computing \cite{KimbleWu1986, LIGO2019, Casacio2021, Madsen2022, Nehra2022, Stokowski2023, Kawasaki2025}.

Three performance metrics -- high gain, broad bandwidth, and low noise figure determine OPA utility in many practical settings. High gain boosts signals above noise floors and compensates for optical losses. Broad bandwidth enables amplification of ultrafast or wavelength-division-multiplexed signals. Low noise figure ensures that the signal-to-noise-ratio is not degraded beyond what is required by the laws of quantum mechanics.

Despite these advantages, integrating OPAs on photonic chips has proven challenging.
Integrated photonics promises to revolutionize optical systems by miniaturizing components and systems from square-meters to square-centimeter scales~\cite{BUTT2025}. For OPAs in particular, this vision of integration poses challenges due to the power consumption of the associated pump lasers.
Figure~\ref{fig_intro}a summarizes the key metrics for a practical integrated OPA:  performance metrics (high gain, broad bandwidth, low noise) need to be achieved within integration constraints (low power, on-chip multiplexing of pump and signals, small footprint). Recent advances have demonstrated high gain, broad gain bandwidth, single-mode operation, and low noise figure on platforms like silicon \cite{Foster2006, Liu2010, Kuyken2011, Wang2012} silicon nitride \cite{Ooi2017, Ye:2021, Riemensberger2022, Ayan:23, Qu2023, Zhao2025},  gallium phosphide \cite{Kuznetsov2025}, and annealed proton-exchanged \cite{Sohler1980, Serkland1995}, mechanically polished \cite{Umeki:11, Kashiwazaki2021}, dry-etched \cite{Kazama:21}, and thin-film lithium niobate \cite{Jankowski:22, Ledezma:22, Stokowski2023, Li25, Chen2025}. However, these continuous-wave OPAs often required watts of optical power provided by off-chip fiber amplifiers to achieve appreciable ($>10$~dB) gain, far exceeding what typical integrated lasers can output.

OPA gain scales exponentially with the product of effective nonlinearity $g$ and device length $L$,
\begin{align} \label{eq_G_OPA}
    G_\text{PSA} =  e^{2gL}, \\
    G_\text{PIA} = \cosh^2{gL},
\end{align} 
where $G_\text{PSA}$ and $G_\text{PIA}$ are the phase-sensitive and phase-insensitive gains, respectively. 
The effective nonlinearity depends on the intrinsic material nonlinearity, the confinement and overlap of the different beams, and pump power. For $\chi^{(2)}$ and  $\chi^{(3)}$ systems the effective nonlinearity scales with normalized efficiency and nonlinear coefficient $\eta_0, \gamma_0$, and pump power $P$ as $ g_{\chi^{(2)}} =\sqrt{\eta_0 P}, \ g_{\chi^{(3)}} = \gamma_0 P$, respectively. Since achieving high gain requires large $gL$, OPAs typically demand substantial pump power. However, practical constraints limit both parameters:
integrated lasers typically provide $10\text{s}-100$s of milliwatts, and device lengths face limits from losses, fabrication variations\cite{Santandrea:19, JieZhao:2023}, bandwidth reduction\cite{Jankowski_2021}, and footprint constraints.

We choose thin-film lithium niobate  for its large $\chi^{(2)}$ nonlinearity and tight optical confinement, yielding an effective nonlinearity orders of magnitude larger than many other platforms \cite{Jankowski_2021}. The platform's low propagation losses\cite{Loncar17,Khalatpour2025} across its wide transparency window ($400$~nm$-5000$~nm) and tight bend radii enable dense integration. 

A $\chi^{(2)}$ OPA requires an optical pump at the second harmonic (SH), i.e., at twice the optical frequency we wish to amplify. Yet generating this pump from the fundamental frequency (FH) telecom lasers is essential for leveraging mature, high-power, and low noise integrated lasers established by the telecommunications industry \cite{Shimizu2023}. Starting with an FH laser also preserves phase coherence for homodyne detection in phase-sensitive applications such as quantum-enhanced sensing and information processing \cite{Kashiwazaki2023, LIGO2019, Stokowski2023, Kawasaki2025}. Generating the SH on chip introduces an additional nonlinear stage whose efficiency must be taken into account. The FH-to-SH conversion efficiency is
\begin{equation} \label{eq_normal_SHG}
\eta_\text{SHG} = \tanh^2{gL}.
\end{equation}
Finite SHG efficiency reduces the SH pump available to the OPA, tightening the power budget for integrated devices.

Here we address the above limitations with a second-harmonic-resonant design that enhances both SHG efficiency and effective pump power for the OPA. Our TFLN device significantly improves all the integration metrics in Figure \ref{fig_intro}a, demonstrating a low-power integrated OPA suitable for practical deployment.

\section{Second Harmonic Resonant OPA Design}\label{sec_design}

\begin{figure*}[ht!]
  \centering
  \includegraphics[width=\linewidth]{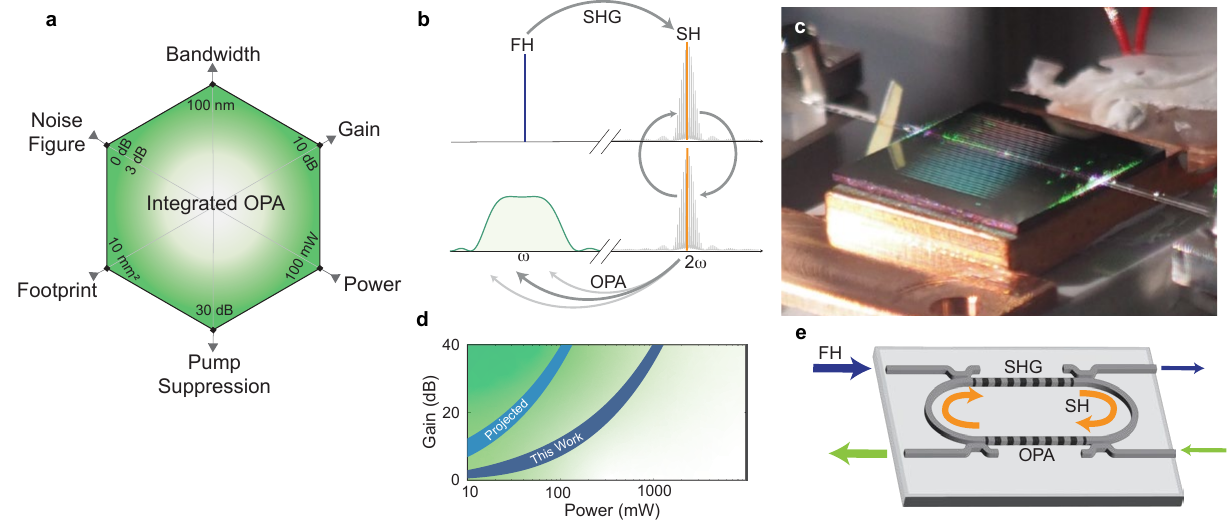} 
\caption{\label{fig_intro} 
(a) Important metrics for an on-chip OPA applications. Representative values are listed - exact metric requirements vary with application. (b) Energy flow in second-harmonic-resonant OPA. Energy flows from the fundamental pump to the broadband amplified signal via the resonant second harmonic pump. (c) Image of fabricated photonic chip atop copper holder. (d) Simulated OPA gain as a function of input power, for parameters measured in this work ($\gamma = 0.3$, $\eta = 2000\frac{\%}{\text{Wcm}^2}$, $L = 6$ mm) and also for improved parameters ($\gamma = 0.3$, $\eta = 4000\frac{\%}{\text{Wcm}^2}$, $L = 1.2$ cm). The top of each curve represents phase-sensitive gain, while bottom represents phase-insensitive gain (Equation (\ref{eq_G_OPA})). Shaded green represents desired integrated OPAs with low power and high gain. (e) Chip implementation of second-harmonic-resonant OPA. All couplers shown are dichroic couplers that couple nearly all light around the fundamental frequency while almost no light at the second harmonic.
} 
\end{figure*}

Our second-harmonic-resonant design simultaneously enhances SHG efficiency and OPA gain by resonating the SH pump while maintaining single-pass operation for the signal and idler. Unlike typical signal-resonant and fully-resonant systems, which achieve ultralow-power nonlinear interactions but limit the bandwidth \cite{McKenna2022, YunZhao:23}, our system combines resonant pump enhancement with broadband traveling-wave amplification.

As shown in Figure \ref{fig_intro}b,e, the FH telecom pump generates resonant SH pump, which in turn pumps an OPA. Dichroic couplers couple nearly 100$\%$ of light near the fundamental frequency (FH pump, signal, idler) but negligible light at the second harmonic, ensuring single-pass signal amplification and effective pump-signal multiplexing. Because the second harmonic is generated within the resonator, there is no impedance matching requirement as in typical resonators \cite{liscidini2019,McKenna2022,xiao2025}. This simplifies device operation, and leads to more robust, scalable, and adaptable systems by removing the need for a optical-power-dependent critical coupling parameter to be engineered into the device. 

The steady-state circulating second harmonic power, $P_\text{SH}$, just after the SHG section satisfies
\begin{align} \label{eq_PCD_SHG}
    P_\text{SH} &= P_{0}\frac{\eta_\text{SHG}(P_\text{SH})}{\gamma},~~~~~\text{where}  \\
    \eta_\text{SHG} &\approx 1-e^{-2\sqrt{\eta_0 P_\text{SH}}L}, \label{eq_eta_PCD_SHG}
\end{align}
where $P_0$ is the input FH power, $\eta_\text{SHG}$ is the FH-to-SH conversion fraction, $L$ is the length of SHG section (assumed to be the same as the OPA section), and $\gamma$ is the SH round-trip loss (this relation holds in the low loss limit; see Extended Data for details). Equation (\ref{eq_PCD_SHG}) highlights the two-fold benefit: both SH power and conversion efficiency are resonantly enhanced. 

Combining equations (\ref{eq_PCD_SHG}) and (\ref{eq_G_OPA}) yields the expressions for phase-sensitive and phase-insensitive gain: 
\begin{align} \label{eq_G_SHROPA}
    G_\text{SHROPA, PSA} = e^{2 L\sqrt{\eta_0 \frac{\eta_\text{SHG}}{\gamma}P_0 }}, \\
    G_\text{SHROPA, PIA} = \cosh^{2}{ L \sqrt{\eta_0 \frac{\eta_\text{SHG}}{\gamma}P_0 }}.
\end{align}
The gain as a function of power is plotted in Figure \ref{fig_intro}d, for both the parameters achieved in this work and the projected parameters given the fabrication capabilities currently demonstrated. In both cases, high gain is achieved at low powers.

\section{Efficient Second Harmonic Generation}\label{secSHG}

\begin{figure}[h!]
  \centering
  \includegraphics[width=\linewidth]{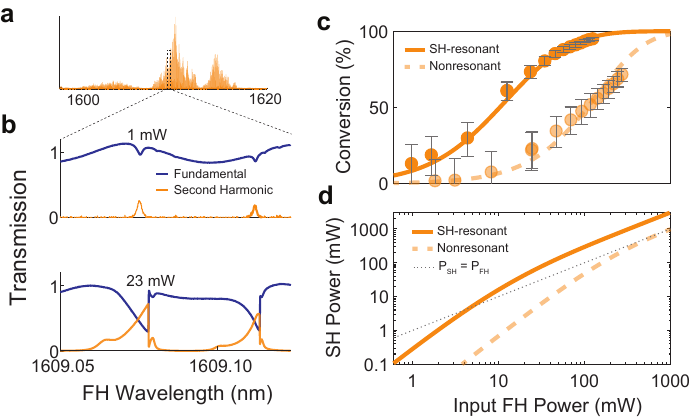} 
\caption{\label{fig_shg} 
Resonant Second Harmonic Generation. (a) SHG transfer function, showing hundreds of SH resonances across 10s of nm. (b) Transmission of FH pump (blue) and leakage of resonant second harmonic (orange, scaled to match FH pump), for two different on-chip input powers. (c) Conversion efficiency from input to second harmonic, measured by depletion of the FH pump. Dots are measured data while lines are model for $\eta_0 = 2000\frac{\%}{\text{Wcm}^2}$ and $\gamma \approx 0.3$. (d) Second harmonic pump power to OPA, for both resonant and nonresonant cases and the same parameters given above. Dotted line shows when the second harmonic power in the OPA is equal to the input fundamental power to the system.
} 
\end{figure}

We observe efficient second-harmonic generation in fabricated devices. Many narrow SH resonances appear as we tune the laser across the phase-matching bandwidth, as shown in Figure \ref{fig_shg}a. Looking at a narrow span of wavelengths that contains only a few resonances, we observe peaks in the generated SH and corresponding dips in the FH transmission (Figure \ref{fig_shg}b). At milliwatt-level powers, we see Lorentzian-squared SH peaks with quality factors of $10^6$, corresponding to a SH round trip loss of $\gamma = 0.3$. The dips in the FH pump transmission are due to conversion to the SH, and are consistent with a normalized efficiency of $\eta_0 = 2000 \frac{\%}{\text{Wcm}^2}$ for this device with $L = 6$ mm nonlinear length. At higher powers, FH depletion deepens and thermally induced resonance shifts broaden the SH lineshapes, increasing tolerance to pump detuning and expanding the usable wavelength range.

Figure \ref{fig_shg}c compares the enhanced conversion efficiency of resonant versus nonresonant SHG. Nonresonant SHG was measured in a diagnostic waveguide through the same nonlinear region as the second-harmonic-resonant waveguide. 
At 124 mW input power, the nonresonant device converts 54\% of the power to the SH, while the resonant device achieved up to 95\% conversion.
Moreover, resonant recirculation increases the total SH pump power in the amplifier by $6\times $ larger in the SH-resonant device compared to the nonresonant device, as shown in Figure \ref{fig_shg}d. This enhancement is most pronounced at low powers ($\frac{4}{\gamma^2}$), and approaches the resonant build-up limit $1/\gamma$ as the conversion efficiency tends to unity.

Further improvements are readily achievable. Attaining the simulated nonlinearity $\eta_0 = 4000 \frac{\%}{\text{Wcm}^2}$ by improved poling, a modest improvement in the finesse (similar to that achieved on another device on the same chip), and extending the nonlinear length to $L=12$ mm would reduce power requirements for SH generation by another order of magnitude. The device performance we project with these modifications are shown as the projected curve in Figure~\ref{fig_intro}d.

\section{Low-Power Optical Parametric Amplification}\label{secOPA}

\begin{figure*}[ht!]
  \centering
  \includegraphics[width=0.9\linewidth]{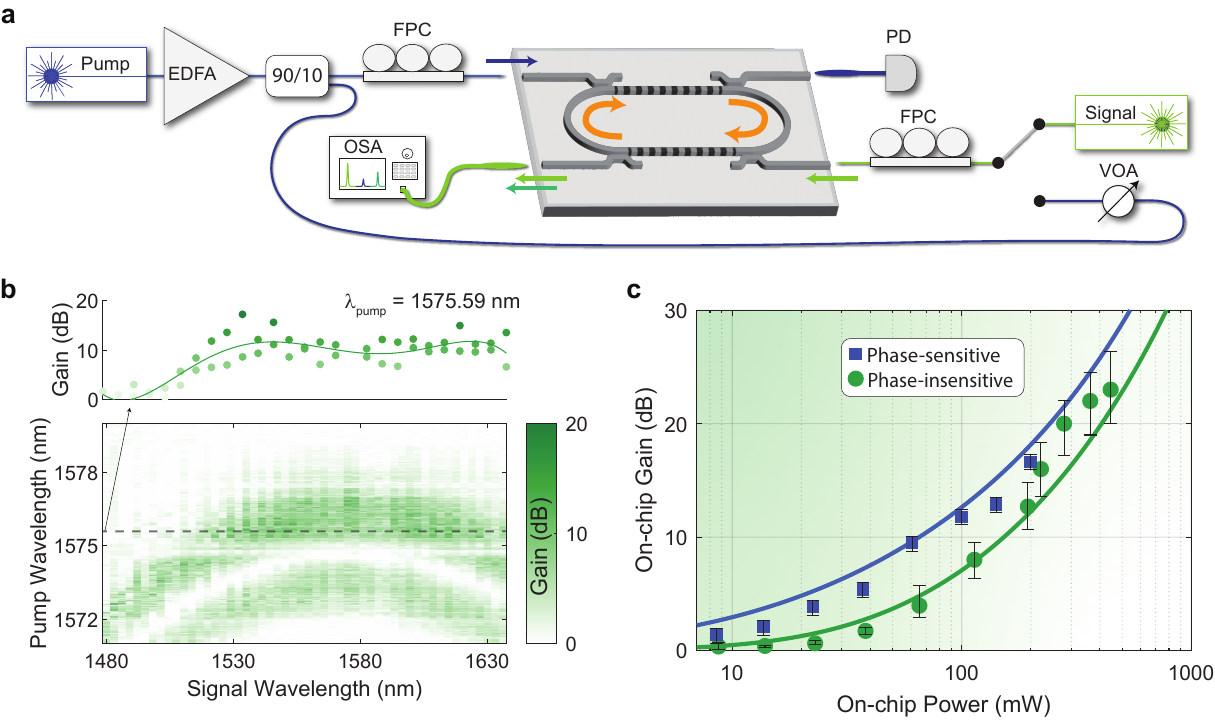}
\caption{\label{fig_opa} (a) OPA Measurement Setup. Fundamental pump laser is amplified by an erbium-doped fiber amplifier (EDFA), sent through a fiber polarization controller (FPC), and coupled by lensed fiber onto the chip, where it generates the resonant second harmonic pump. A nondegenerate signal laser is input on the other side of the chip for phase-insensitive amplification measurements. For phase-sensitive measurements, a tap of the fundamental pump laser is input as degenerate seed instead. Output light is collected by multimode fiber and measured by an optical spectrum analyzer (OSA). (b) Lower Plot: OPA on-off gain spectrum for on-chip FH pump power of 225 mW. The gain at each signal wavelength is measured while the FH pump wavelength is tuned. Upper Plot: Gain spectrum for fixed pump wavelength at 1575.59 nm on a single SH resonance. Points are extracted from one horizontal slice of the lower plot, and the line is to guide the eye. (c) On-chip net gain as a function of on-chip FH pump power, for signal wavelengths around 1590 nm. Green markers represent phase-insensitive amplification measurements and blue markers represent phase-sensitive ones. Error bars on phase-insensitive amplification measurements show the minimum and maximum gain to accurately show the gain ripple due to chip-facet reflections as discussed in section \ref{sec_methods_OPA}. Curves are calculated based on the SHG performance of section \ref{secSHG}.  
} 
\end{figure*}

We characterize the performance of the OPA by coupling the FH pump and the signal to be amplified from opposite facets of the chip (Fig.~\ref{fig_opa}a).
Tuning the wavelengths of both lasers results in the OPA gain spectrum shown in Figure \ref{fig_opa}b for one device with nonlinear length $L=9$ mm and input power 225 mW. The on-off gain is calculated by dividing the amplified signal transmission by the baseline signal transmission with the pump wavelength detuned from phase matching at 1580 nm.
The curved tuning shape and the measured bandwidth are consistent with the simulated dispersion (Section \ref{sec_methods_OPA_Gain_spectrum}). The horizontal and curved gaps in gain are due to the SHG and OPA phase matching for this device.
As the pump wavelength is tuned, more than 12 dB on-chip net gain is observed across the entire 160 nm signal laser tuning range.  The top inset in Figure \ref{fig_opa}b plots a single slice of the gain spectrum and shows broadband $> 110$~nm gain for a single pump wavelength.

Figure \ref{fig_opa}(c) depicts net gain vs FH pump power. We measure over 17 dB phase-sensitive gain for degenerate wavelengths around 1576 nm and 12 dB of phase-insensitive gain for signal wavelengths around 1590 nm at less than 200 mW input power. To our knowledge, this is the first continuous-wave, integrated OPA to achieve substantial ($>10$ dB) broadband gain at less than 300 mW pump power.
At 450 mW, the phase-insensitive gain increases to 23 dB.  Even accounting for the 4.8 dB/facet fiber-chip coupling loss (which could be improved to below 0.6 dB \cite{Hu:21}), the device achieves fiber-chip-fiber net gain at pump powers below 200 mW.

\section{Broadband Quantum-limited Amplification}\label{secNF}

\begin{figure}[h!]
  \centering
  \includegraphics[width=\linewidth]{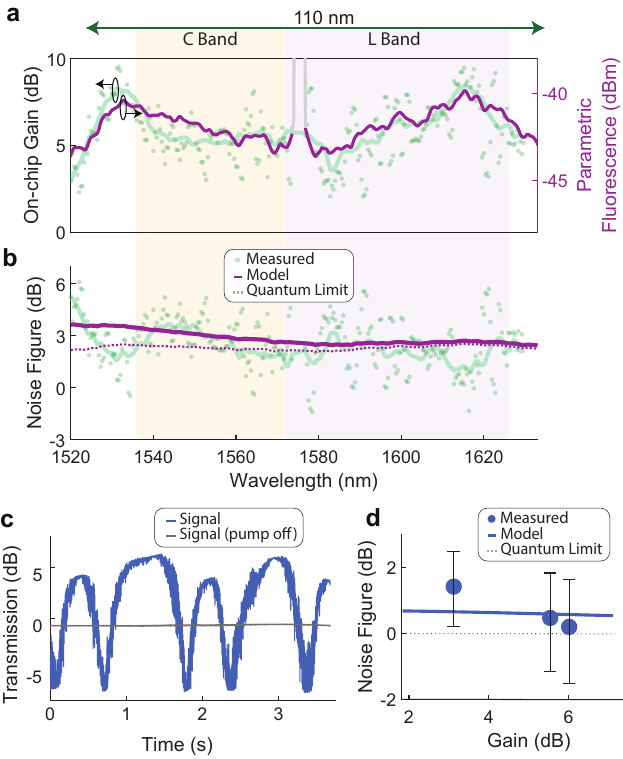} 
\caption{\label{fig_NF} 
OPA Noise Figure Measurements. (a) Signal gain for one FH pump wavelength around 1575.31 nm, alongside the spontaneous parametric fluorescence spectrum with 2 nm resolution setting on the optical spectrum analyzer. (b) Phase-insensitive noise figure, as a function of signal wavelength. Solid purple line represents the expected noise figure based on measured losses and the SPF spectrum of (a). Dotted line represents the quantum-limit of noise figure based on the SPF spectrum of (a). (c) Degenerate amplification as phase drifts in time. (d) Phase-sensitive noise figure based on SPF level and amplification. Points are measured datasets from SPF and gain for three adjacent SH pump modes, as in (c). Curve represents expected noise figure based on measured losses. Dotted line represents the 0 dB quantum limit.
} 
\end{figure}

We achieve low-power operation together with broad bandwidth and a low noise figure.
We quantify the bandwidth and noise by tuning the FH pump wavelength to SH resonance and measuring the gain and spontaneous parametric fluorescence (SPF) (Figure \ref{fig_NF}a).
The 3-dB amplification bandwidth is 110 nm based on the spontaneous parametric fluorescence spectrum, in good agreement with the measured gain spectrum and nearly three times the bandwidth of a typical EDFA. 
Even larger gain bandwidths are possible: a slightly shorter $L = 6$ mm device has over 150 nm of bandwidth while dispersion-engineered thin-film lithium niobate OPAs promise flat gain bandwidths in excess of 340 nm \cite{Jankowski_2021, Ledezma:22}. 
The variance in the gain data is due to weak reflections from the end-facets of the chip which impart a few-dB ripple in the gain spectrum, as discussed in the SI Section \ref{sec_methods_OPA}. 
Comparing the on-chip gain and fluorescence yields the on-chip noise figure shown in Figure \ref{fig_NF}b \cite{OPANFtheory}. The noise figure is quantum limited across most of the 1520 - 1630 nm band. The increase in noise at around 1520 nm is due to the wavelength dependence of the pump-signal multiplexers as discussed in Extended Data Section \ref{sec_methods_NF}.

Figure \ref{fig_NF}c shows the phase-sensitive transmission of a degenerate signal. Without any phase locking, the relative phase between the FH pump and degenerate seed drifts in time with temperature fluctuations and vibrations along their fiber paths, resulting in the observed oscillatory amplification and deamplification cycles. 
The extracted phase-sensitive noise figure is shown in \ref{fig_NF}d for three adjacent resonances and approaches just 0.5 dB, close to the quantum-limit of 0 dB for phase-sensitive amplifiers and successfully circumventing the 3-dB noise penalty inherent to phase-insensitive amplification\cite{Ye:2021, Kazama:21}.  
We are able to measure degenerate amplification despite a strong FH pump input thanks to the 26 dB of FH pump suppression provided by our dichroic couplers. Even more suppression is within reach - another device on the same chip achieved 52 dB, and there is little cost to adding more filters to the design due to the minimal effect of the evanescent couplers on the SH transmission.

\section{Discussion}\label{sec_discussion}

We have demonstrated a second-harmonic-resonant architecture that delivers broadband parametric gain at very low input powers. Using the large quadratic nonlinearity of TFLN, an FH pump is efficiently doubled to SH inside an SH-resonant device. The intracavity SH field then pumps a single-pass OPA to provide significant near-quantum-limited amplification over $110$~nm.

Our key advance is to decouple pump power build-up from signal bandwidth: the FH pump is doubled to SH inside an SH-resonant section, and the intracavity SH field then drives a single-pass OPA. This resonant-pump/single-pass-signal configuration boosts the effective pump by the cavity build-up while preserving the broadband, dispersion-set gain of a traveling-wave amplifier. Because the SH is generated intracavity, no power-dependent impedance matching is required, simplifying operation and enabling robust pump–signal multiplexing via dichroic couplers.

Current limitations are primarily engineering.  Weak end-facet reflections introduce few-dB gain ripple which is exacerbated at higher pump power. This can be mitigated with AR coating or better coupler design. More efficient edge couplers also benefit fiber applications - adiabatic couplers can reduce fiber-chip coupling loss to below $0.6$~dB\cite{Hu:21}. Further filtering of the FH and SH pumps can suppress them by $>50$~dB, while immediate improvements in poling quality, resonator quality factor, and device design can reduce required pump power further by an order of magnitude, enabling significant gain at 10s of milliwatt pump powers.

Beyond these refinements, our approach generalizes across wavelength bands accessible in TFLN or any other quadratically nonlinear material, offering a practical route to integrated, broadband, low-noise amplification, and squeezed-light generation. Thanks to reduced power requirements, co-integration with current chip-scale lasers, possible by existing hybrid photonic technologies, will enable a fully integrated OPA subsystem suitable for quantum and classical applications.

\section{Methods}\label{sec_methods}


\subsection{Comparison with Literature}

\begin{figure}[h!]
  \centering
  \includegraphics[width=\linewidth]{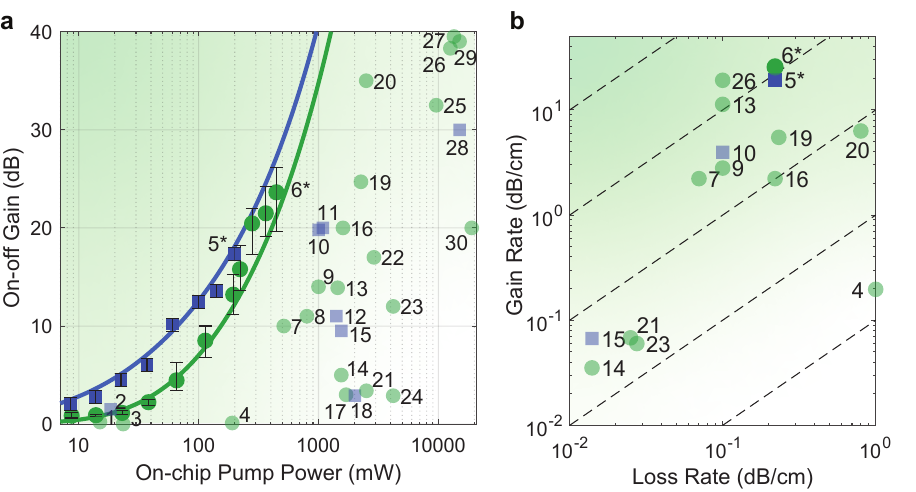}
\caption{\label{fig_litreview} 
(a) Gain versus pump power for chip-scale OPAs in the literature (see Table \ref{litreview_table}). Green circles represent phase-insensitive amplification measurements and blue squares phase-sensitive amplification. Darker points correspond to data from this work. Curves are simulations using the same parameters as Section \ref{secSHG}. (b) Gain rate (at max reported power) vs loss rate of chip-scale OPAs. Dashed lines are lines of constant nonlinearity-to-loss ratio. Those references that did not include loss rate information could not be plotted.
} 
\end{figure}

\begin{table*}[]
\begin{tabular}{lllllllllll}
Label & Ref.                     & Platform & Type & P (W) & G (dB)  & L (cm) & g (dB/cm) & $\alpha$ (dB/m) & $g/\alpha $    & On-Chip Filters \\
\hline
1     & \cite{Wang2012}          & Si       & PIA  & \textbf{0.02}  & 0.21   & 0.8    & 0.27      &            &   & No              \\
2     & \cite{Stokowski2023}     & TFLN     & PSA  & \textbf{0.02}  & 1.5    & 1      & 1.5       &            &   & Yes             \\
3     & \cite{Ooi2017}           & SiN      & PIA  & \textbf{0.02}  & 0.01   & 0.7    & 0.02      &            &   & No              \\
4     & \cite{Li25}              & TFLN     & PIA  & \textbf{0.19}  & 0.12   & 0.6    & 0.2       & 100             & 0.2    & Yes             \\
\textbf{5*}     &\textbf{This Work}                & TFLN     & PSA  & \textbf{0.2}   & \textbf{17.36}  & 0.9    & 19.29     & 22              & 87.69  & Yes             \\
\textbf{6*}     &\textbf{This Work}                & TFLN     & PIA  & \textbf{0.45}  & \textbf{23.63}  & 0.9    & 26.26     & 22              & 119.35 & Yes             \\
7     & \cite{Kashiwazaki2021}   & LN       & PIA  & \textbf{0.51}  & \textbf{10}     & 4.5    & 2.22      & 7               & 31.75  & No              \\
8     & \cite{Cestier:12}        & GaInP    & PIA  & \textbf{0.8}   & \textbf{11}     & 0.15   & 73.33     &            &   & No              \\
9     & \cite{Kishimoto:16}      & LN       & PIA  & 1     & \textbf{14}     & 5      & 2.8       & 10              & 28     & No              \\
10    & \cite{Kishimoto:16}      & LN       & PSA  & 1     & \textbf{19.8}   & 5      & 3.96      & 10              & 39.6   & No              \\
11    & \cite{Kashiwazaki2023}   & LN       & PSA  & 1.09  & \textbf{20}     & 4.5    & 4.44      &            &   & No              \\
12    & \cite{Umeki:11}          & LN       & PSA  & 1.41  & \textbf{11}     & 5      & 2.2       &            &   & No              \\
13    & \cite{Chen2025}          & LN       & PIA  & 1.45  & \textbf{13.9}   & 1.23   & 11.3      & 10              & 113.01 & No              \\
14    & \cite{Ye:2021}           & SiN      & PIA  & 1.55  & 5      & 142    & 0.04      & 1.4             & 2.52   & No              \\
15    & \cite{Ye:2021}           & SiN      & PSA  & 1.55  & 9.5    & 142    & 0.07      & 1.4             & 4.78   & No              \\
16    & \cite{Shimizu2023}       & LN       & PIA  & 1.6   & \textbf{20}     & 9      & 2.22      & 22              & 10.1   & No              \\
17    & \cite{Ayan:23}           & SiN      & PIA  & 1.7   & 3      & 200    & 0.02      &            &   & No              \\
18    & \cite{Serkland1995}      & LN       & PSA  & 2.01  & 2.9    & 1      & 2.9       &            &   & No              \\
19    & \cite{Kazama:21}         & LN       & PIA  & 2.26  & \textbf{24.7}   & 4.5    & 5.49      & 23.26           & 23.6   & No              \\
20    & \cite{Kuznetsov2025}     & GaP      & PIA  & 2.5   & \textbf{35}     & 5.55   & 6.31      & 80              & 7.88   & No              \\
21    & \cite{Zhao2025}          & SiN      & PIA  & 2.51  & 3.4    & 50     & 0.07      & 2.5             & 2.72   & No              \\
22    & \cite{Shimizu2024}       & LN       & PIA  & 2.9   & \textbf{17}     &   &      &            &   & No              \\
23    & \cite{Riemensberger2022} & SiN      & PIA  & 4.2   & \textbf{12}     & 200    & 0.06      & 2.75            & 2.18   & No              \\
24    & \cite{Foster2006}        & Si       & PIA  & 4.2   & 2.9    & 1.7    & 1.71      &            &   & No              \\
25    & \cite{Lamont:08}         & As2S3    & PIA  & 9.6   & \textbf{32.5}   & 6.5    & 5         &            &   & No              \\
26    & \cite{Sua:18}            & LN       & PIA  & 12.6  & \textbf{38.3}   & 2      & 19.15     & 10              & 191.5  & No              \\
27    & \cite{Kuyken2011}        & Si       & PIA  & 13.5  & \textbf{39.5}   & 2      & 19.75     &            &   & No              \\
28    & \cite{Ledezma:22}        & TFLN     & PSA  & 15.04 & \textbf{30}     & 0.6    & 50        &            &   & No              \\
29    & \cite{Ledezma:22}        & TFLN     & PIA  & 15.04 & \textbf{39}     & 0.6    & 65        &            &   & No              \\
30    & \cite{Liu2010}           & Si       & PIA  & 19    & \textbf{20}     & 0.4    & 50        &            &   & No              \\
31    & \cite{Qu2023}            & SiN      & PIA  & 180   & \textbf{24}     & 2      & 12        &            &   & No              \\
32    & \cite{Sohler1980}        & LN       & PIA  & 200   & 2.43   & 3.2    & 0.76      &            &   & No             
\end{tabular}
\caption{ \label{litreview_table} Performance of broadband chip-scale OPAs, in order of ascending pump power. Fundamental pump powers less than 1 W as well as gains greater than 10 dB are shown in bold. }
\end{table*}

A comparison of integrated OPAs is shown in Table \ref{litreview_table}, sorted in order of increasing pump power. We have attempted to accurately extract on-chip pump power and on-off gain from each reference. For those references that performed SHG to generate the pump (all $\chi^{(2)}$ references), we report the FH pump power. We report the peak power of pulsed measurements.

Figure \ref{fig_litreview}a plots gain versus power as in Figure \ref{fig_opa}(c) but now includes the references of Table \ref{litreview_table}. 
Figure \ref{fig_litreview}b plots the nonlinearity and gain rates. The nonlinearity-to-loss ratio typically scales with power, but we achieved a nonlinearity-to-loss ratio well over 100 even at the low on-chip powers of this work. The nonlinearity-to-loss ratio is important for loss-sensitive applications such as the generation and detection of squeezed light. 

\subsection{OPA Gain Spectrum}\label{sec_methods_OPA_Gain_spectrum}

\begin{figure}[h!]
  \centering
  \includegraphics[width=\linewidth]{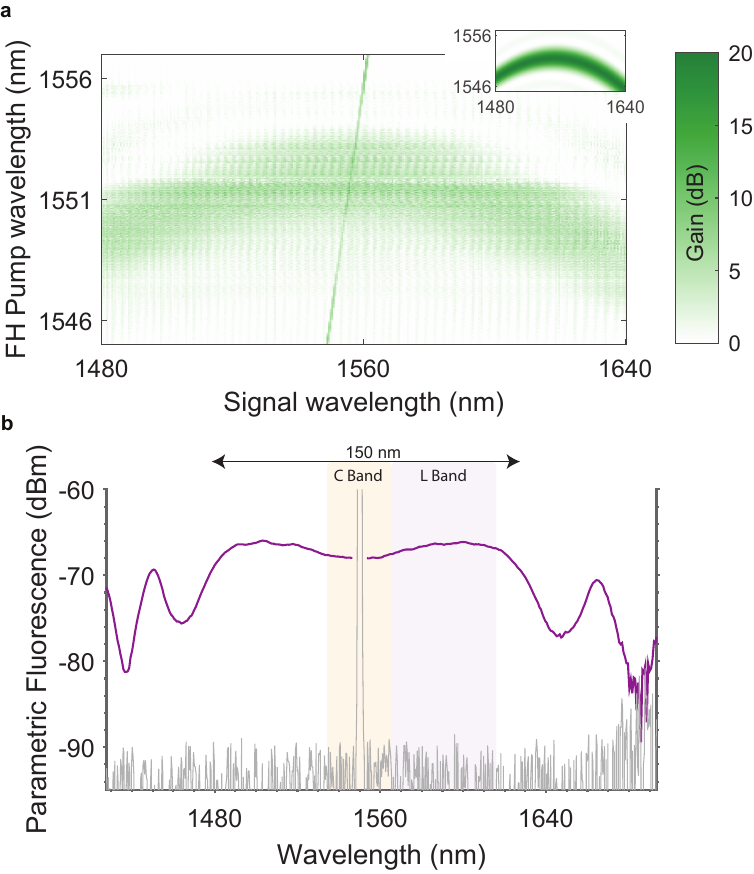} 
\caption{\label{fig_dev2heatmap} (a) OPA gain spectrum for a $L=6$ mm device. Inset: OPA gain spectrum as predicted by simulated dispersion. (b) Spontaneous parametric fluorescence spectrum verifying bandwidth exceeds 150 nm.
} 
\end{figure}

We measured the OPA gain spectrum of Figure \ref{fig_opa}b by recording signal transmission while the FH pump wavelength was swept across the SHG phase matching bandwidth. The generation of resonant SH pump led to amplification of those signal wavelengths that were phase matched.
Figure \ref{fig_dev2heatmap}a depicts a densely sampled gain spectrum with signal wavelengths separated by 0.5 nm for a different $L=6$ mm device. A small amount of FH pump leakage is transmitted through the dichroic filters and appears as a thin line through the center of the plot. The curved shape and bandwidth are consistent with the simulated gain spectrum shown in the inset, based on the theoretical dispersion.
The SPF spectrum for pump wavelength 1551.1 nm has 150 nm 3-dB bandwidth as shown in Figure \ref{fig_dev2heatmap}b.

\subsection{OPA Measurements}\label{sec_methods_OPA}

\begin{figure}[h!]
  \centering
  \includegraphics[width=\linewidth]{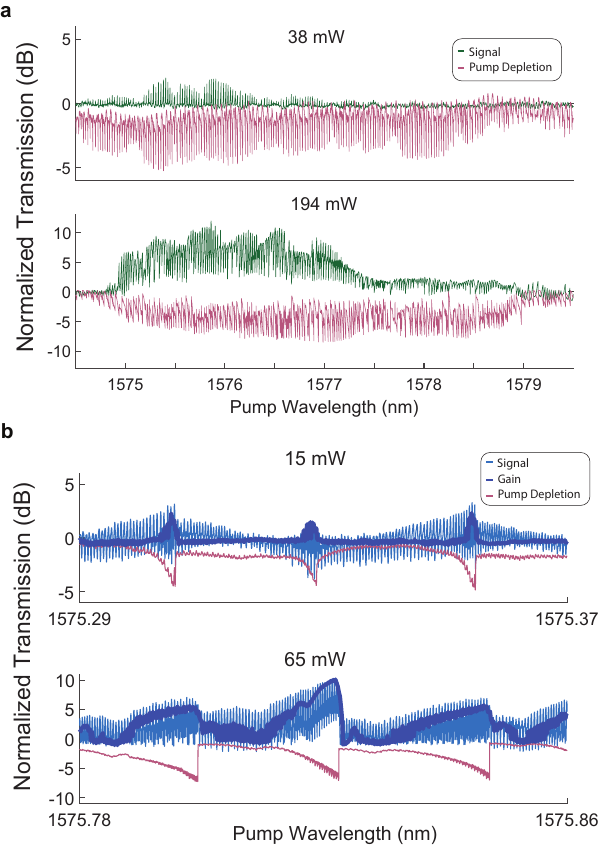} 
\caption{\label{fig_Gextract} 
    Gain measurements, at various on-chip input powers. (a) Phase-insensitive gain from OPA and corresponding FH pump depletion from SHG for nondegenerate signal wavelength 1590 nm and on-chip pump powers 38 mW and 194 mW. 
    (b) Phase-sensitive gain from OPA and corresponding FH pump depletion from SHG for degenerate signal wavelengths for on-chip pump powers of 15 mW and 65 mW.
} 
\end{figure}

We measure nondegenerate OPA gain on and off phase matching by recording signal transmission while the pump wavelength is swept for fixed signal wavelength as in Figure \ref{fig_Gextract}. Dips in the FH pump transmission align with peaks in the signal transmission as expected - the SH resonances both deplete the FH pump and amplify the signal. At higher powers, we observe more pump depletion and more gain.
To account for the slight gain ripple due to on-chip reflections (see Extended Data), we repeat these measurements multiple times for slightly different signal wavelengths and report the mean peak gain, with error bars for the min and max peak gains as shown in Figure \ref{fig_opa}.

We perform similar measurements for degenerate OPA. 
Amplification and deamplification cycles occur as relative phase changes with wavelength due to the fiber path length difference between input and seed as shown in Figure \ref{fig_Gextract}b. The magnitude of amplification and deamplification increases as the wavelength approaches the (thermally broadened) resonance, as visible by the correlation of gain magnitude with input depletion. However, the low electronic bandwidth of the photodiode used in these measurements (Yokogawa AQ6374 OSA $\approx 
10$ kHz analog output bandwidth) smoothed out the very fast deamplification cycles at high powers. We extract the true gain $G$ from these measurements from the average signal transmission over one cycle:
\begin{multline} \label{eq_G_DOPA_extract_avg}
    G = (\langle G_\text{observed}\rangle - \frac{P_\text{l,G}}{P_\text{s,0}}) \\ + \Re\bigg[\sqrt{(\langle G_\text{observed}\rangle - \frac{P_\text{l,G}}{P_\text{s,0}})^2 -1}
\bigg]
\end{multline}
where $\langle G_\text{observed}\rangle$ is the smoothed transmission over one cycle and $\frac{P_\text{l,G}}{P_\text{s,0}}$ captures the effect of the (small) amount of FH pump leakage $P_\text{l,G}$ that can linearly interfere with the input signal $P_\text{s,0}$.


\subsection{Noise Figure Measurements} \label{sec_methods_NF}

The noise figure of an optical parametric amplifier depends on the net gain $G_\text{net}$ and generated spontaneous parametric fluorescence (SPF) $\rho_\text{ASE,out}$. We extract on-chip net gain by measuring the on-off gain $G_\text{on/off}$, dichroic coupler input and output losses $1-\eta_1, \ \ 1- \eta_2$, and propagation losses $1-\eta_\text{prop}$
\begin{equation}
    G_\text{net} = \eta_1\eta_2 \eta_\text{prop} G_\text{on/off}.
\end{equation}
The SPF level is directly measured on the OSA after calibrating for chip-fiber coupling loss.
The noise figures for our phase-insensitive and phase-sensitive amplification measurements are calculated by 
\begin{align} \label{eq_NF_PIA}
NF_\text{PIA}  = \frac{2 \frac{\rho_\text{ASE, out}}{h\nu}}{G_\text{net} } + \frac{1}{   G_\text{net}}
 \\
\label{eq_NF_PSA}
NF_\text{PSA} = \frac{G_\text{PSA,on/off}}{G_\text{PIA,on/off}} \bigg(  \frac{ \frac{\rho_\text{ASE}}{ h\nu}}{G_\text{net} }  + \frac{1}{G_\text{net}} \bigg)
\end{align}

\begin{figure}[h!]
  \centering
  \includegraphics[width=\linewidth]{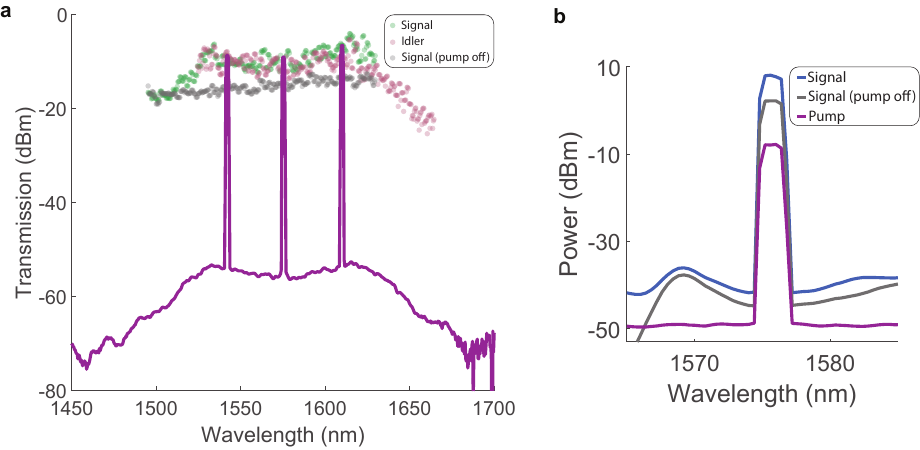} 
\caption{\label{fig_NFextract} 
    Extracting noise figure from SPF spectrum. (a) Signal and generated idler measured with the max-hold technique, as well as spectrum for a single signal wavelength input. (b) Degenerate amplification spectra. SPF level is extracted from off-degenerate reading, and the gain is extracted not from this single scan but instead by measuring the amplification/deamplification cycles in time as in Figure \ref{fig_NF}(c).
} 
\end{figure}

For noise figure measurements only, we insert a tunable bandpass filter in the FH pump path prior to coupling to the chip in order to filter any EDFA amplified spontaneous emission so that we can effectively measure the SPF.
We tune the FH pump onto an SH resonance and leave it there without any active feedback. We find the resonances are stable for at least tens of minutes, particularly at higher powers where the resonances are thermally broadened, and therefore wavelength-cavity locking is not necessary for our measurements.
For phase-insensitive noise figure measurements, we measure on-off gain using the max-hold technique \cite{Riemensberger2022, Kuznetsov2025} and slowly sweep signal wavelength while continuously updating the maximum of the OSA spectrum.
For phase-sensitive noise figure measurements, we measure on-off gain by observing degenerate signal amplification and deamplification in time and applying Equation (\ref{eq_G_DOPA_extract_avg}). Figure \ref{fig_NFextract}b shows example OSA spectra with the on-chip SPF level around -50 dBm for 2 nm OSA resolution.

\subsection{Dichroic Coupler Characterization}\label{sec_methods_DC}

\begin{figure}[h!]
  \centering
  \includegraphics[width=\linewidth]{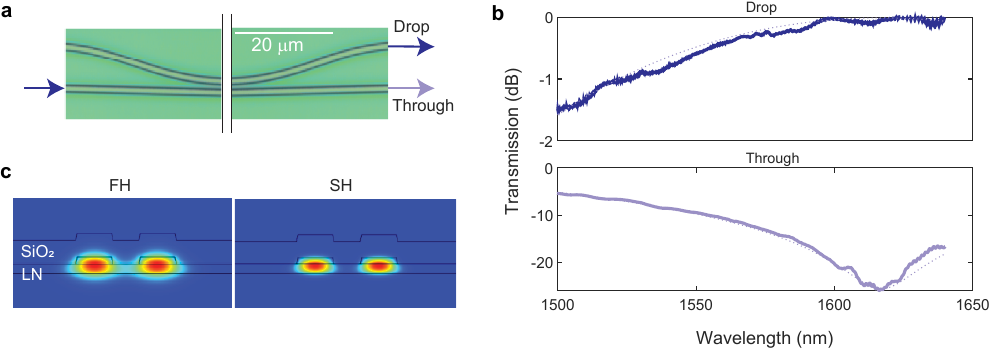} 
\caption{\label{fig_DC} 
(a) Microscope Image of fabricated dichroic coupler (0.5 mm straight coupling region is cropped out). (b) Transmission spectra of dichroic coupler drop (top) and through (bottom) ports. (c) Simulated symmetric mode profiles of fundamental and second harmonic in coupler region.
} 
\end{figure}

Our dichroic couplers serve to transmit signals with low loss and suppress the FH and SH pumps.
Figure \ref{fig_DC}a shows their physical implementation as directional couplers. The length (500 $\mathrm{\mu m}$) and gap (1.11 $\mathrm{\mu m}$) are chosen so that nearly all FH light is coupled to the drop port while nearly all SH light transmits to the through port. This dichroic behavior occurs due to the different mode sizes at FH and SH wavelengths shown in Figure \ref{fig_DC}c - the FH mode is shared between the coupled waveguides while the SH mode is not. 

Figure \ref{fig_DC}b plots the FH through and drop port transmission spectra for a single dichroic coupler, with the rapid ripple from chip-facet reflections smoothed out (see Extended Data for details). 
The signal light is coupled with less than one dB of coupling loss across the entire OPA amplification bandwidth of 1520 - 1630 nm, and as small as 0.04 dB coupling loss at the peak transmission around 1617 nm. We estimate that the entire one-dB bandwidth of the couplers exceeds 200 nm.
If even greater coupling bandwidths are required, adiabatic couplers can be used \cite{Guo:16, Guo2022, Li25}.

The through-port extinction is important to suppress the FH pump from leaking into the OPA section, where it can interfere with the signal. We observe up to 26 dB peak extinction from a single dichroic, and 13 dB at our operating wavelength of 1575 nm.  The two dichroic couplers separating the SHG and OPA sections of our device combine to double the total FH pump suppression, and the efficient SHG conversion contributes up to 13 dB additional suppression by depleting the FH pump. Future designs could readily incorporate more dichroic couplers to achieve additional FH pump suppression. 

We estimate that between 0.1\% and 1\% of SH light couples to the drop port in our dichroic couplers, based on Equation (\ref{eq_PCD_SHG}) and pump depletion measurements. This low outcoupling helps keep the SH resonance high finesse and also suppresses the SH pump from the OPA output.

\subsection{Fabrication}\label{sec_methods_fab}

\begin{figure}[h!]
  \centering
  \includegraphics[width=\linewidth]{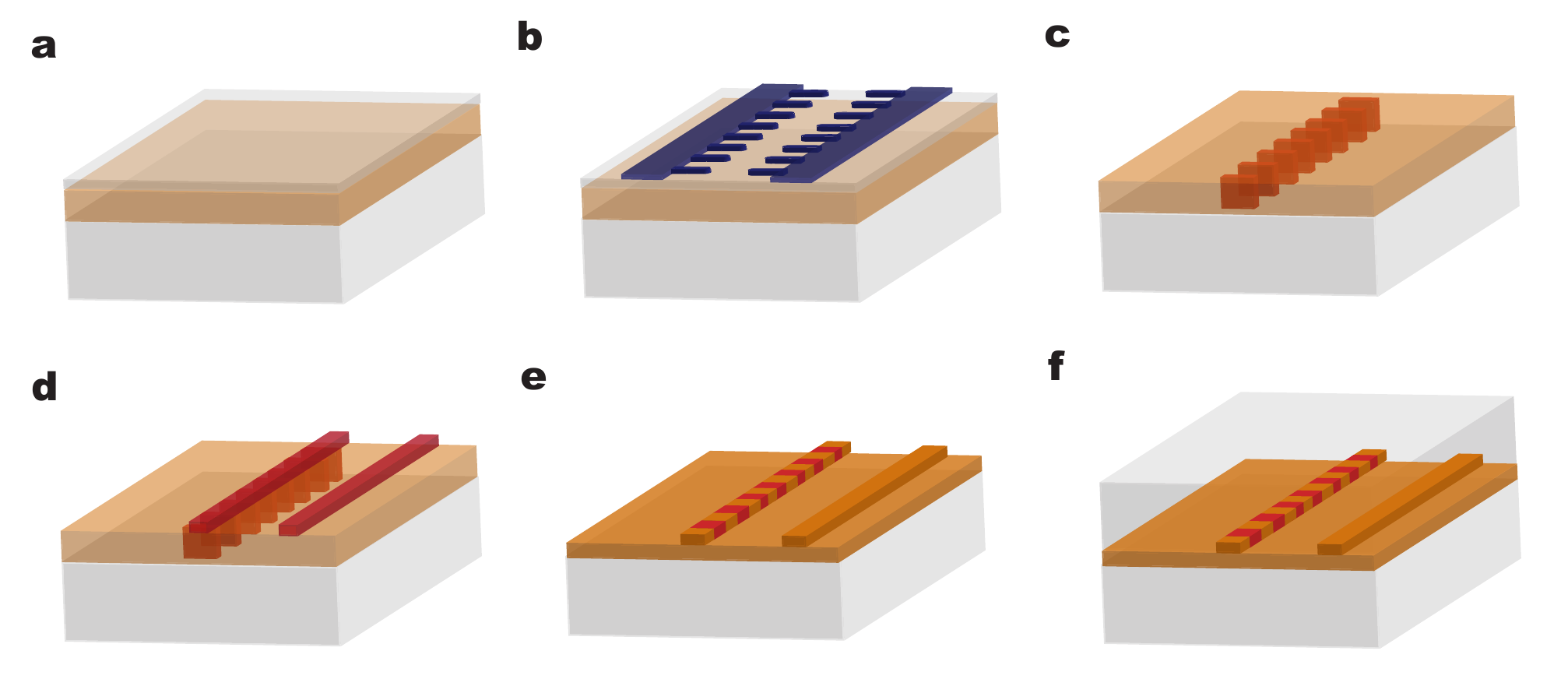} 
\caption{\label{fig_FAB} 
Fabrication procedure. 
\textbf{a},~ Deposit 100 nm of SiO\textsubscript{2} onto thin-film lithium niobate on insulator chip.
\textbf{b},~Use electron beam lithography to pattern and then liftoff 100 nm-thick aluminum electrodes where the poling periods are adaptively designed to compensate for thickness variations across the film.
\textbf{c},~Apply high voltage pulses to periodically pole the LN and then remove the electrodes.
\textbf{d},~Pattern HSQ mask using electron beam lithography for waveguide patterning.
\textbf{e},~Argon ion mill 300 nm of LN to etch the waveguides and then acid cleaning to produce patterned LN waveguides. 
\textbf{f},~Deposit 700 nm of SiO\textsubscript{2} for cladding using HDPCVD.
} 
\end{figure}

We start with an X-cut LNOI (NanoLN) chip with a 700 nm-thick LN layer to fabricate the OPA. First, we thin the chip to 500 nm using Ar ion milling. At this step, we pattern alignment marks using a hydrogen silsesquioxane (HSQ) mask and electron beam lithography to enable aligned writes in later steps. We then map the chip thickness using reflectometry (Filmetrics F40) to characterize film thickness variations.

Next, we deposit 100 nm of SiO\textsubscript{2} using a high-density plasma-enhanced chemical vapor deposition (Plasma-Therm Versaline HDPCVD) system and anneal it at 500 \textdegree C for 8 hours. On top of the SiO\textsubscript{2}, we pattern aluminum electrodes for periodic poling using electron beam lithography (EBPG 5200+) and a bi-layer polymethyl methacrylate (PMMA) liftoff process. The poling periods are locally adjusted to compensate for thickness variations that induce phase mismatch\cite{Chen2024, Xin2025}. 

We periodically pole the regions corresponding to the SHG and OPA sections (two distinct periodically poled regions) by applying high-voltage pulses to the electrodes, with a peak voltage of 600 V and pulse durations of 1 ms–1 ms–15 ms (rise, hold, fall times). After poling, we remove the Al electrodes using 25$\%$ TMAH and strip the 100 nm SiO\textsubscript{2} layer with buffered oxide etchant (BOE).

We then pattern an HSQ mask using electron beam lithography and etch the LN via Ar ion milling to define the device. The waveguide geometry follows previous work\cite{McKenna2022, Stokowski2023}, with a ridge waveguide width of 1.2 µm, a height of 500 nm, and an etch depth of 300 nm. Using the HDPCVD system, we deposit a 750 nm SiO\textsubscript{2} cladding layer. Finally, we anneal the chip in an oxygen environment at 500 \textdegree C for 8 hours. The chip is diced using a DISCO DFL7340 laser saw afterwards.

\section{Extended Data}

The extended data contains additional information regarding SH-resonant SHG calculations, propagation losses, fiber-chip coupling, transmission and facet reflections, gain and noise figure, pump-signal multiplexing, and saturation effects.


\subsection{SH-resonant OPA Derivations}\label{sec_methods_PCD_deriv}

We solve for the resonant second harmonic power in SH-resonant SHG by solving self-consistently for the power after one round trip pass. In the low-loss limit the SH is approximately constant throughout the SHG section 
\begin{equation}
        P_\text{SH} \approx P_\text{SH}(1-\gamma) + \eta_\text{SHG} P_0
\end{equation}
where $P_\text{SH}$ is the intracavity SH power just after the SHG section, $\gamma$ is the round-trip loss (including any effective loss from OPA conversion), $\eta_\text{SHG}$ is the SHG conversion efficiency, and $P_0$ is the input FH pump power. In contrast with linear resonators described by fields linearly interfering based on fixed coupling strengths, this nonlinear resonator is much simpler to describe in terms of power and conversion.
Rearranging, we arrive at equation \ref{eq_PCD_SHG}:
\begin{align} \label{eq_deriv_PCD_SHG_approx}
    P_\text{SH} &= P_{0}\frac{\eta_\text{SHG}}{\gamma},~~~~~\text{where}  \\
    \eta_\text{SHG} &\approx 1-e^{-2\sqrt{\eta_0 P_\text{SH}}L}, 
\end{align}
where the expression for $\eta_\text{SHG}$ comes from integrating the nonlinear coupled-wave equations in the limit of constant SH power. 
This is a simple and powerful equation for describing SH-resonant OPA since it directly relates the measured SHG pump depletion to the expected phase-sensitive OPA gain by (for equal SHG and OPA nonlinear lengths) $\eta_\text{SHG} = 1 - \frac{1}{G}$. 

We also derive a more robust equation that is valid even for large round trip losses. Assuming about half the loss comes from the SHG section and using the exact solution for lossless SHG evolution \cite{Li:95}, we revise equation \ref{eq_deriv_PCD_SHG_approx} to get
\begin{align} \label{eq_deriv_PCD_SHG_lumped}
    P_\text{SH, after} &= \left(1-\frac{\gamma}{2}\right) \times \nonumber\\&P_\text{SH}\left[L, \eta_0, P_0, \left(1-\frac{\gamma}{2}\right) P_\text{SH, after}\right],\\
    \eta_\text{SHG} &= \frac{P_0 - P_\text{FH}(L, \eta_0, P_0, (1-\frac{\gamma}{2})P_\text{SH, after})}{P_0}
\end{align}
where $P_\text{SH, after}$ is the SH power just after the SHG section (an important clarification since the SH power is no longer near constant throughout the resonator), $P_\text{SH}(z,\eta_0, P_0, P_\text{SH,in})$ and $P_\text{FH}(z,\eta_0, P_0, P_\text{SH,in})$ describes the lossless SH and FH evolution during the SHG section\cite{Li:95}, and $P_\text{SH,in}$ is the SH power at the input to the SHG section.

\begin{figure}[h!]
  \centering
  \includegraphics[width=\linewidth]{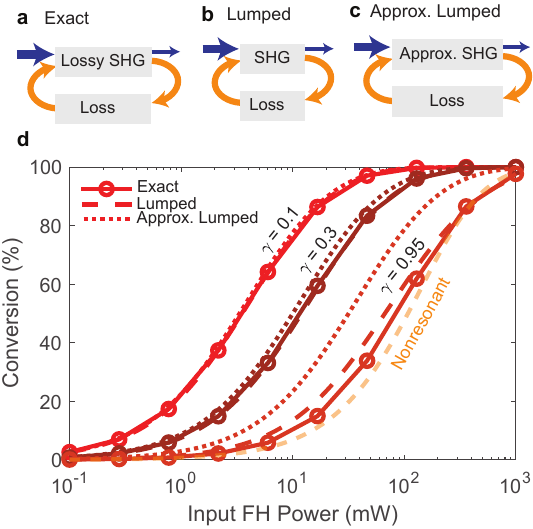} 
\caption{\label{fig_sim_model} 
Numerically simulations and simple model for conversion efficiency for SH-resonant SHG.
} 
\end{figure}

We verify the validity of equations (\ref{eq_deriv_PCD_SHG_approx}) and (\ref{eq_deriv_PCD_SHG_lumped}) by comparing them with full numerical simulations that directly solve the lossy nonlinear coupled wave equations for many round trips to find the steady-state SH power distribution.
As shown in Figure \ref{fig_sim_model}d, the approximate model of equation \ref{eq_deriv_PCD_SHG_approx} is valid up until fairly large losses ($\gamma \approx 0.5$) and the more precise model of equation \ref{eq_deriv_PCD_SHG_lumped} is robust for essentially all loss values. At very high losses, the solution approaches that of nonresonant SHG since the resonant enhancement disappears with loss.

\subsection{Low-power SH Resonance Lineshape} \label{sec_SI_lorentzian}
At low powers, SH-resonant SHG produces Lorentzian-squared resonances that scale quadratically with input power and finesse.
The self-consistent field equation for the SH after one round trip is
\begin{equation}
    A_\text{SH} = A_\text{SH}\sqrt{1-\gamma}e^{i\phi} +  \sqrt{ P_0 \eta_\text{SHG}},
\end{equation}
where $A_\text{SH}$ is the intracavity SH field amplitude, $\phi$ is the round-trip phase, and $\sqrt{ P_0 \eta_\text{SHG}}$ is the added SH amplitude from the SHG section.

At low input powers, the conversion efficiency becomes $\eta_\text{SHG} \approx  2 \sqrt{\eta_0 P_\text{SH}} L$. 
At low losses, the round-trip field transmission becomes $\sqrt{1-\gamma} \approx 1 - \frac{\gamma}{2}$. Close to resonance, the round-trip phase can be approximated as $e^{i\phi} \approx 1 + i\phi$. Under these approximations, the circulating SH field and power are
\begin{align} 
    A_\text{SH} &\approx A_\text{SH}(1-\frac{\gamma}{2})(1+i\phi) +  \sqrt{P_0 L A_\text{SH} \sqrt \eta_0}, \notag \\
    & \text{which results in } A_\text{SH} \approx \frac{\sqrt\eta_0 P_0 L}{(\frac{\gamma}{2} - i\phi)^2}, \notag  \\
    & \text{or, }P_\text{SH} = |A_\text{SH}|^2 =  \frac{\eta_0 P_0^2L^2}{(\frac{\gamma^2}{4} + \phi^2)^2}.\label{eq_lorentz_derv}
\end{align}
Note that on resonance $\phi = 0$, the resonant enhancement of SH-resonant SHG compared to nonresonant SHG is $\frac{4}{\gamma^2}$ in this low-power limit. As discussed in the main text, at high powers it saturates to $\frac{1}{\gamma}$. At such high powers, the wings of the lineshape increase by more than the saturated peak and results in a power-broadened lineshape similar to those observed in atomic absorption \cite{LEVINE2012}.

To explicitly compare our low-power SH-resonant lineshape with the typical Lorentzian lineshape of a linear resonator, we divide by round-trip time $\tau$ to define the usual total loss rate $\kappa = \frac{\gamma}{\tau}$ and detuning $\Delta = \frac{\phi}{\tau}$. We define the effective extrinsic coupling rate as $\kappa_e = \frac{\eta_\text{SHG}}{\tau} = \frac{\eta_0 P_0 L^2}{\tau} $ and then express the power build-up of equation \ref{eq_lorentz_derv} 
\begin{align} \label{eq_lorentz_squared}
    P_\text{SH} \approx P_0 \bigg(\frac{ \kappa_e}{(\frac{\kappa}{2})^2 + \Delta^2} \bigg)^2,
\end{align}
which describes the square of a Lorentzian resonance. 
In contrast with linear resonators, whose impedance matching requirements form a tradeoff between input-coupling efficiency and resonant enhancement\cite{siegman86, McKenna2022}, this nonlinear coupling\cite{liscidini2019, xiao2025} to SH resonator has no such impedance matching since the extrinsic coupling rate $\kappa_e$ does not contribute to the total loss rate $\kappa$. 


\subsection{SH-resonant OPA Scaling with Length}

The pump power in SH-resonant OPA depends on round-trip pump propagation losses 
\begin{equation}
    P_\text{SH} \approx \frac{P_0 \eta_\text{SHG}}{\gamma} = \frac{P_0 \eta_\text{SHG}}{1-e^{-2\alpha_\text{SH}L_c}} \approx \frac{P_0\eta_\text{SHG}}{2\alpha_\text{SH}L_c}, 
\end{equation}
where $L_c$ represents the full cavity length and $\alpha_\text{SH}$ is the SH field propagation loss rate, and the last approximation was in the limit of small round trip losses. We can approximate $L_c \approx 2L$ if we assume most of the length of the cavity comes from the SHG and OPA nonlinear sections.
Assuming sufficiently efficient SHG $\eta_\text{SHG} \approx 1$, the length-dependence of the SH-resonant OPA equations becomes
\begin{align}
    G = e^{2\frac{g}{\sqrt\gamma}L} = e^{2g \frac{L}{\sqrt{2\alpha_\text{SH}L_c}}} \notag\\ 
    \approx   e^{2g \sqrt{\frac{L}{4\alpha_\text{SH}}}}.
\end{align}
The exponential scaling with square-root of nonlinear length suggests that long, low-finesse SH resonators are yield higher gains compared to short, high-finesse resonators with the same SH loss rate. As mentioned in the text, the maximum length is often fixed by bandwidth, fabrication variation, or footprint constraints. In quantum applications, required nonlinearity-to-loss ratio is another factor that could drive SH-resonant OPA designs towards smaller, higher-finesse resonators. 

Note that the gain of nonresonant OPAs can also be limited by pump propagation losses.
The OPA gains of equation \ref{eq_G_OPA} are modified in the presence of pump propagation losses by $L \rightarrow L_\text{eff}$, where the effective length captures the diminishing gain as the pump accrues losses: 
\begin{equation}
    L_\text{eff} = \frac{1-e^{-\alpha_\text{SH}L}}{\alpha_\text{SH}}.
\end{equation}
At small lengths, $L_\text{eff} \approx L$ and the effect of losses is negligible. At longer lengths, however, $L_\text{eff} \rightarrow \frac{1}{\alpha_\text{SH}}$ and the losses limit effective length and therefore gain.

\subsection{FH and SH Propagation Losses}\label{sec_methods_Qs}

The SH resonances exhibit squared-Lorentzian lineshapes at low powers (see Figure \ref{fig_shg}b, Section \ref{sec_SI_lorentzian}) with quality factors around $1\times10^6$ and round-trip losses of $\gamma \approx 0.3$ in the 18 mm cavities.
The SH propagation loss rate is therefore bounded by 
\begin{equation}
    \alpha_\text{SH} < 80 \text{ dB/m}
\end{equation}
by associating all losses with intrinsic propagation loss.

The FH intrinsic losses are small enough that we cannot accurately measure them by direct transmission measurements. 
Instead, we bound the FH propagation and dichroic coupler losses from a resonant measurement on an identical chip. We measure quality factors up to $1.93\times 10^6$ in an FH resonator that includes a 7 mm periodically poled region and two dichroic couplers. 
Diagnostic racetrack resonators can display quality factors up to $4\times 10^6$, so we can bound the loss contributions from propagation loss in the poled region and in the couplers
\begin{align}
   10 \text{ dB/m} < \alpha_\text{poled} < 34 \text{ dB/m} \\
   -0.083 \text{ dB} < \gamma_\text{coupler} < -0.013 \text{ dB},
\end{align}
where the lower and upper bounds come from attributing the excess measured loss to either the poled region or the couplers. The upper bound on dichroic coupler drop-port transmission is from the measured peak through-port extinction mentioned in section \ref{sec_methods_DC}.
The total intrinsic loss for the signal transmitted through our $L = 9 $ mm OPA device with two dichroic couplers is therefore bounded by $-0.29 \pm 0.04 \text{ dB}$.


\subsection{Fiber-Chip Coupling}\label{sec_methods_fiberchipcoupling}

\begin{figure}[h!]
  \centering
  \includegraphics[width=\linewidth]{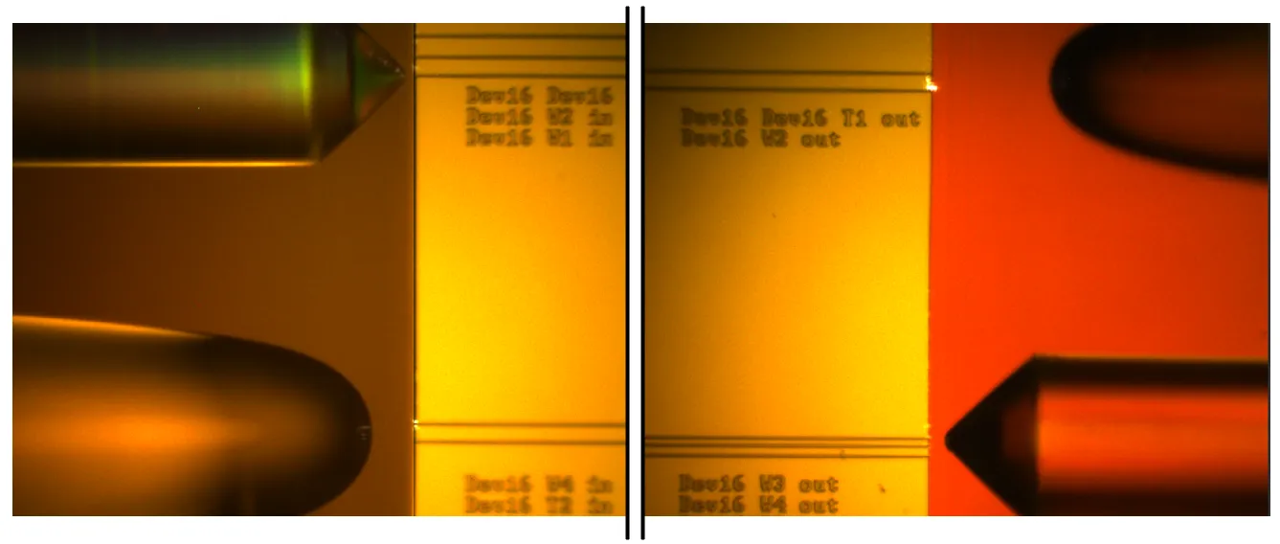} 
\caption{\label{fig_SMFMMF} 
Microscope image of fibers coupling light onto and off of the chip. Lensed single mode fiber (top left and bottom right) couple light onto the chip, while lensed multi-mode fiber (bottom left and top right) collect light from the chip.
} 
\end{figure}

For OPA measurements where multiple input and output beams are required, we use lensed single mode fibers (SMF) to couple light into the single mode waveguides and lensed multi-mode fibers (MMF) to collect output light. Figure \ref{fig_SMFMMF} shows the pairs of SMF and MMF fibers aligned on opposite sides of the chip. We measure -4.8 dB/facet fiber-chip coupling loss with the SMF fibers. The large core size of the MMFs makes them largely insensitive to alignment. Fibers are glued to a V-groove chip with 250 $\mathrm{\mu m}$ pitch. 

For all other measurements where only a single input beam is required, we switch to only input and collect light with the SMFs. The collected fundamental is separated from the second harmonic by a dichroic mirror, and then the fundamental and second harmonic are attenuated by variable optical attenuators and then focused onto InGaAs and Si avalanche photodiodes for detection.

\subsection{Transmission and Facet Reflections}\label{sec_methods_trans}

\begin{figure}[h!]
  \centering
  \includegraphics[width=\linewidth]{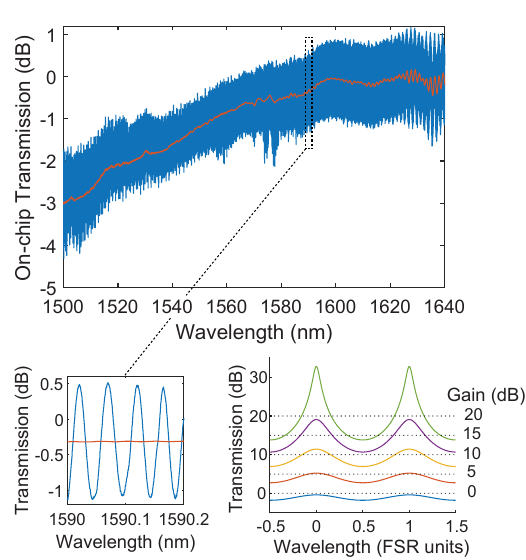} 
\caption{\label{fig_transmission} 
(a) On-chip signal transmission spectrum through device. Inset: zoom in on sinusoidal ripples due to end-facet reflections on either side of the chip, creating a weak standing-wave cavity. (c) Simulated amplified signal transmission in the presence of facet reflections, for gains ranging from 0 to 20 dB. The gain ripples increase close to threshold, where net gain approaches net round-trip loss.
} 
\end{figure}
Figure \ref{fig_transmission}a depicts the on-chip transmission spectrum of light through the signal path - consisting of two dichroic couplers with a long straight section in the middle - normalized to the transmission spectrum of a straight diagnostic waveguide. The fiber-chip-fiber transmission can be obtained by adding in the -4.8 dB/facet fiber-chip coupling loss and the -0.29 dB intrinsic propagation loss. 
The transmission decreases at shorter wavelengths due to the dichroic coupler transmission discussed in Section \ref{sec_methods_DC}. 

The sinusoidal ripples in the transmission are due to end-facet reflections creating a weak standing-wave cavity with transmission 
\begin{equation}
    T_\text{tot} = \frac{1}{1 +   \frac{4R\sqrt{\mathcal{G}}}{(1-R\sqrt{\mathcal{G}})^2} \sin^2{\frac{\phi}{2}}},
\end{equation}
where $R$ is the end facet reflectivity, $\mathcal{G}$ is the on-chip net gain, $\phi$ is the phase accumulated in one round trip \cite{siegman86}. Fitting the linear transmission data gives a reflectivity of about $R = 9\%$ per facet. 
These ripples lead to a slightly wavelength-dependent transmission and, as a secondary effect, a wavelength-dependent gain. 
We account for these ripples when measuring phase-insensitive gain by sampling many (8) seed wavelengths within a single ripple period, thereby obtaining the minimum, maximum, and mean gain in the presence of the ripple. Figure \ref{fig_opa}c depicts this variance by plotting data points at the mean gain, with error bars stretching to the minimum and maximum gain observed, for each power level. Note that the spontaneous parametric fluorescence measurements of Figure \ref{fig_NF} were recorded on an optical spectrum analyzer with 2 nm resolution and therefore averaged over these gain ripples, unlike the gain measurements which fluctuate by a few dB with the ripple. 

It was noted in \cite{Kuznetsov2025} that the ripples affect bidirectional amplifiers such as laser gain media much more than they do unidirectional optical parametric amplifiers. Still, at very high pump powers the ripples can lead to optical parametric oscillation.
These ripples can be mitigated by AR coatings, index matching fluids, and adiabatic or angled end facet couplers. 
Furthermore, because they only arise when coupling light both onto and off of the chip, they should be greatly reduced in systems with monolithically integrated lasers or detectors.


\subsection{Phase-sensitive Gain - Data Analysis}

In this section we model the wavelength-dependent phase-sensitive amplification of the degenerate OPA measurements to arrive at equation \ref{eq_G_DOPA_extract_avg}. The FH pump and degenerate seed beams are generated from a single input laser incident on a beamsplitter, travel different paths in fiber, and then are coupled onto the chip. The FH pump generates the SH pump, whose phase determines the amplification/deamplification quadratures for the seed light. The seed mixes with a small amount of FH pump leakage that is not entirely suppressed by the filters and then coupled into the OPA.

The phases of all the beams at the start of the OPA section depend on the effective path lengths $L_i$ and propagation constants $\beta_i$:
\begin{align*}
\phi_\text{seed} &= \beta_\text{FH} L_\text{seed}, \\
\phi_\text{SH} &= 2(\beta_\text{FH}L_\text{pump}) + \beta_\text{SH}L_\text{SHG}, \\
\phi_\text{leakage} &= (\beta_\text{FH}L_\text{pump}) + \beta_\text{FH}L_\text{SHG},
\end{align*}
where $L_\text{SHG}$ is the small path length between the SHG and OPA sections and $L_\text{seed}, L_\text{pump}$ capture the large mostly-fiber path lengths traveled by seed and pump beams after the beamsplitter. Since the phase of the SH determines what quadratures are amplified and deamplified, we define the phases
\begin{align}
\phi  &\equiv  \phi_\text{SH} - 2\phi_\text{seed} \\&= 2\beta_\text{FH}(L_\text{pump} - L_\text{seed}) + \beta_\text{SH}L_\text{SHG}, \notag  \\
\theta &\equiv \phi_\text{SH} - 2\phi_\text{leakage} \\&= (\beta_\text{SH} - 2\beta_\text{FH})L_\text{SHG}. \notag
\end{align}
As we tune wavelength across a single SH resonance, the relative phase $\phi$ between SH and seed changes rapidly but the relative phase $\theta$ between SH and leakage is essentially constant due to the relatively small path length $L_\text{SHG} \ll L_\text{pump} - L_\text{seed}$.

The total seed input to the OPA is the linear interference between the injected seed and the (largely suppressed) FH pump leakage. With respect to the amplification and deamplification quadratures $X_s$ and $Y_s$, the total seed quadratures are given by
\begin{align*}
X_s &= \sqrt{P_{s,0}} \cos\phi + \sqrt{P_l} \cos\theta, \\
Y_s &= \sqrt{P_{s,0}} \sin\phi + \sqrt{P_l} \sin\theta.
\end{align*}
The OPA amplifies one quadrature and deamplifies the other, resulting in output power 
\begin{align}
P &= G X^2_s + \frac{1}{G}Y^2_s \quad \notag\\
  &= G\!\left(P_{s,0} \cos^2\phi + P_l \cos^2\theta + 2\sqrt{P_{s,0}P_l}\cos\theta\cos\phi \right) \notag \\
  &\quad + \frac{1}{G}\!\left(P_{s,0} \sin^2\phi + P_l \sin^2\theta + 2\sqrt{P_{s,0}P_l}\sin\theta\sin\phi \right) \notag\\
&= P_{s,0}\!\left(G\cos^2\phi + \frac{1}{G}\sin^2\phi\right) \notag \\
 &\quad + P_l\!\left(G \cos^2\theta + \frac{1}{G}\sin^2\theta\right) \notag \\
&\quad + 2\sqrt{P_{s,0}P_l}\!\left(G\cos\theta\cos\phi  + \frac{1}{G}\sin\theta\sin\phi \right).  \label{eq_P_DOPA_extract}
\end{align}
The first term is the desired OPA phase dependence that varies rapidly with wavelength, the second term is the amplified leakage that varies slowly with wavelength, and the last term is the undesired linear seed-leakage interference that varies rapidly with wavelength. In our measurements, the amplified leakage term was very small compared to the amplified signal ($\approx 10$ dB below). However, slight asymmetric oscillations due to the seed-leakage interference term could be observed for certain gains/seed powers, and also the limited bandwidth of the analog output channel of the optical spectrum analyzer partly smoothed over the most rapid oscillations with wavelength in equation. \ref{eq_P_DOPA_extract}. 

We extract the actual gain $G$ from measured transmission $G_\text{observed}$ by using a moving average filter with period equal to the oscillation period. The smoothing cancels out any linear interference terms and leaves just the desired average gain term incoherently added with the amplified leakage term:
\begin{align}
\langle G_\text{observed} \rangle &= \frac{\langle P \rangle}{P_{s,0}} \notag \\
&= \tfrac{1}{2}\!\left(G + \frac{1}{G}\right)  \notag \\
  & \quad + \frac{P_l}{P_{s,0}}\!\left(G \cos^2\theta + \frac{1}{G}\sin^2\theta\right) \notag \\
&= \tfrac{1}{2}\!\left(G + \frac{1}{G}\right) + \frac{P_{l,G}}{P_{s,0}}, 
\end{align}
where we used the fact that the average over one oscillation of a sinusoid is equal to one half and defined $P_{l,G}$ to be the amplified leakage value for its near-constant phase $\theta$. Solving for $G$:
\begin{align}
G = \left(\langle G_\text{observed}\rangle - \frac{P_{l,G}}{P_{s,0}}\right) 
+ \sqrt{\left(\langle G_\text{observed}\rangle - \frac{P_{l,G}}{P_{s,0}}\right)^2 -1} \notag
\end{align}
which in the high-gain and small-leakage limit simplifies to
\begin{align}
G \approx 2\langle G_\text{observed} \rangle.
\end{align}
In practice, noise on observed gain can lead to negative terms inside the square root around unity gain. To make this equation robust, we restrict it to be real and arrive at equation \ref{eq_G_DOPA_extract_avg}.

\subsection{Noise Figure Derivation}\label{sec_methods_NF_deriv}
The on-chip noise figure of a phase-insensitive amplifier (PIA), assumed to be limited by signal-spontaneous beat noise and shot noise \cite{OPANFtheory}, depends on the phase-insensitive net gain $G_\text{net}$ and measured output spontaneous parametric fluorescence (SPF) level $\rho_\text{ASE,out}$:
\begin{equation*}
    NF_\text{PIA}  = \frac{2 \frac{\rho_\text{ASE, out}}{h\nu}}{G_\text{net} } + \frac{1}{ G_\text{net}}
\end{equation*}
We can directly measure on-chip SPF level $\rho_\text{ASE,out}$ with the OSA (correcting for chip-OSA coupling efficiency). We extract net gain from measurements of on-off gain $G_\text{on/off}$, dichroic coupler input and output losses $1-\eta_1, \ \ 1- \eta_2$, and propagation losses $1-\eta_\text{prop}$
\begin{equation*}
    G_\text{net} = \eta_1\eta_2 \eta_\text{prop} G_\text{on/off}.
    \end{equation*}
The theoretical SPF level should account for the nonlinearity to loss ratio as discussed in \cite{Ye:2021}, but can be simply bounded by assuming half the propagation loss occurs before the amplifier and the other half after. The noise figure equation we use on our data and theory curves are therefore
\begin{align}
NF_\text{PIA, expt} &=  \frac{2 \frac{\rho_\text{ASE, out}}{h\nu}}{\eta_1\eta_2 \eta_\text{prop} G_\text{on/off} }\notag\\ & ~~~ + \frac{1}{ \eta_1\eta_2 \eta_\text{prop} G_\text{on/off}}
 \\
NF_\text{PIA, theory} & <  \frac{2  (\eta_2 (\frac{1+\eta_\text{prop}}{2}) G_\text{on/off} - 1)}{\eta_1\eta_2\eta_\text{prop} G_\text{on/off} } \notag \\ &~~~+ \frac{1}{ \eta_1\eta_2\eta_\text{prop} G_\text{on/off}}
\end{align}

Similarly, the phase-sensitive amplification noise figure is
\begin{align}
    NF_\text{PSA} &=\frac{\frac{G_\text{PSA}}{G_\text{PIA}}(\frac{\rho_\text{ASE}}{h\nu\Delta\nu} + 1)}{G_\text{PSA}} \notag
\\ &= \bigg( \frac{ \rho_\text{ASE}}{G_\text{PIA} h\nu} + \frac{1}{G_\text{PIA}}
 \bigg)
\end{align}
and our experimental and theoretical expressions are
\begin{align}    
NF_\text{PSA, expt} &=  \frac{1}{\eta_1\eta_2\eta_\text{prop} }\frac{G_\text{PSA,on/off}}{G_\text{PIA,on/off}} \notag \\ & \times \bigg(  \frac{ \frac{\rho_\text{ASE}}{ h\nu}}{G_\text{PSA,on/off}} + \frac{1}{G_\text{PSA,on/off}}  \bigg)
\\
NF_\text{PSA, theory} &=  \frac{ \eta_2(\frac{1+\eta_\text{prop}}{2}) (G_\text{PIA,on/off}-1)}{\eta_1\eta_2\eta_\text{prop} G_\text{PIA,on/off} } \notag\\ & + \frac{1}{\eta_1\eta_2\eta_\text{prop}G_\text{PIA,on/off}} ,
\end{align}
where $G_\text{PSA,on/off}$ is the measured on-off gain and $G_\text{PIA,on/off}$ is the  phase-insensitive on/off gain that corresponds to the measured phase-sensitive gain (equation \ref{eq_G_OPA}).

\subsection{Saturation Effects}

Saturation effects are out of the scope of this work and will be reported in a future study. Initial calculations reveal that the saturation behavior of SH-resonant OPA has some beneficial properties compared to nonresonant OPAs, since the SH power is stabilized and enhanced by the cavity.
Furthermore, parasitic sum-frequency generation between the FH pump and signal, which contributes significantly to channel crosstalk in wavelength-division-multiplexed single-stage OPAs, does not occur in this design because the SHG and OPA sections are distinct and so the FH pump and signal do not overlap\cite{Li25}.

\section*{Acknowledgements}         
\vspace{-2mm}
This work was supported in part by the Defense Advanced Research Projects Agency (DARPA) INSPIRED program (HR00112420356). Part of this work was performed at the Stanford Nano Shared Facilities (SNSF) and Stanford Nanofabrication Facility (SNF), supported by the National Science Foundation under award ECCS-2026822. 
We also thank NTT Research for their financial and technical support. D.D. acknowledges support from the NSF GRFP (No. DGE-1656518). H.S. acknowledges support from the Urbanek Family Fellowship. We are grateful for insightful discussions with Professor Joseph Kahn, Dr. Darwin Serkland at Sandia, Dr. Justin Cohen at DARPA, Dr. Geun Ho Ahn, and Mr. Kevin Multani.  
\vspace{-2mm}

\section*{Author Contribution}
D.D. and A.H.S.-N. conceived the idea. D.D. designed the device, with input from T.P. and H.S.. T.P. fabricated the devices, with assistance from D.D., L.Q., S.R., A.H., and J.H.. D.D. performed the experiments and analyzed the data. H.S., T.P., M.M.F., and A.H.S.-N. provided experimental and theoretical support.

\section*{Ethics Declarations}
    \subsection*{Competing Interests}
D.D., T.P., H.S., and A.H.S.-N. are inventors on a patent application that covers methods for achieving quantum advantage in power-constrained photonic sensors. The other authors declare no competing interests.





\bibliography{sn-bibliography}

\end{document}